\documentclass{article}
\usepackage{a4}
\usepackage{epsfig}
\usepackage{amsfonts}
\usepackage{amsmath}

\title{Trigonometric Solutions of the WDVV Equations from Root Systems}
\author{R. Martini \and L.K. Hoevenaars}
\date{January 2003}

\begin{document}
\bibliographystyle{plain}
\maketitle

\begin{abstract}
By introduction of an additional variable and addition of a Weyl invariant correction term to the perturbative prepotential in five-dimensional Seiberg-Witten theory we construct solutions of the WDVV equations of trigonometric type for all crystallographic root systems.
\end{abstract}

\section{Introduction} 
In two-dimensional topological conformal field theory the following remarkable system of third order nonlinear partial differential equations for a function $F$ of $N$ variables emerged
\begin{equation}\label{eq1}
F_i F_1^{-1} F_j = F_j F^{-1}_1 F_i \qquad i,j = 1, \ldots , N
\end{equation}

Here $F_i$ is the matrix
\[
(F_i)_{kl} = \frac{\partial^3 F}{\partial a_i \partial a_k \partial a_l}\;.
\]

Moreover, it is required that $F_1$ is a constant and invertible matrix. Usually this system is called the WDVV equations. Generalizations, not requiring $F_1$ to be constant, have been introduced and studied in the context of four- and five-dimensional $N=2$ supersymmetric gauge theory.

Although extremely difficult to solve in general, this overdetermined system (\ref{eq1}) of nonlinear partial differential equations admits exact solutions. For instance, within the theory of Frobenius manifolds, a substantial class of polynomial solutions has been constructed by Dubrovin \cite{Dub} for any Coxeter group. Furthermore, for any gauge group, perturbative approximations to exact prepotentials in four-dimensional Seiberg-Witten theory satisfy the (generalized) WDVV equations. These solutions are of rational type and may be constructed for any root system (see \cite{Marsh} and \cite{Mart}).\\

In this note we shall construct solutions of system (\ref{eq1}) of trigonometric type for any crystallographic root system. This construction is achieved by the introduction of an additional variable and addition of a Weyl invariant correction term to the perturbative prepotential in five-dimensional Seiberg-Witten theory.

\section{Main result} 
For our convenience by renumbering the variables we may suppose that $F_N$ is constant and invertible instead of $F_1$. More precisely we have the following result

\subsection{Theorem}
Let $R$ be a crystallographic root system in $\mathbb{R}^n$ and $W$ the corresponding Weyl group. Then the function $F$ of $n+1$ variables $a_1, \ldots, a_n, a_{n+1}$ given by
\begin{equation}\label{eq2}
F(a_1, \ldots , a_n, a_{n+1}) = \frac{1}{2} \sum_{\alpha \in R} f((\alpha, a)) + \gamma (\frac{1}{6} a^3_{n+1} + \frac{1}{2} a_{n+1} (a, a))
\end{equation}

satisfies the system (\ref{eq1}) of WDVV equations. Here $(\alpha, a) = \alpha_1 a_1 + \ldots + \alpha_n a_n$ is the standard Euclidean inner product in $\mathbb{R}^n$ and $f$ is the function given by
\begin{equation}\label{eq3}
f(x) = \frac{1}{6} x^3 - \frac{1}{4} Li_3 (e^{-2x}) = \frac{1}{6} x^3 - \frac{1}{4} \sum^{\infty}_{k=1} \frac{e^{-2kx}}{k^3}
\end{equation}
so that
\begin{equation}\label{eq4}
f''' (x) = \mbox{coth} (x).
\end{equation}

$\gamma$ is some fixed constant satisfying $-\gamma^2 = c$ for some fixed number $c$ depending on the root system $R$. For each type of crystallographic root system the value of $c$ is given in table \ref{tab1}.

{\bf Proof}\\
Obviously the matrix $F_{n+1}$ equals $\gamma I$, where $I$ is the identity matrix. So in this case the WDVV conditions reduce to
\[
F_i F_j = F_j F_i
\]
and are automatically satisfied if $i=n+1$ or $j=n+1$. Therefore we may restrict ourselves to $i,j\leq n$ and in order that the expression $F$ satisfies the WDVV equations the expression
\[
\sum^n_{k=1} F_{ilk} F_{kjm} + F_{il, n+1} F_{n+1, jm} = \sum^n_{k=1} F_{ilk} F_{kjm} +\gamma^2 \delta_{il} \delta_{jm}
\]
should be symmetric in $i$ and $j\leq n$. In other words we should have 
\begin{equation}\label{eq5}
\sum^n_{k=1} (F_{ilk} F_{kjm} - F_{jlk} F_{kim}) + \gamma^2 (\delta_{il} \delta_{jm} - \delta_{jl} \delta_{im})= 0
\end{equation}

The left hand side of the equation (\ref{eq5}) above is anti symmetric in $l$ and $m$, thus this condition is equivalent to
\begin{equation}\label{eq6}
\sum^n_{k=1} F_{ilk} F_{kjm} - F_{jlk} F_{kim} - F_{imk} F_{kjl} + F_{jmk} F_{kil} + 2\gamma^2 (\delta_{il} \delta_{jm} - \delta_{jl}\delta_{im}) = 0
\end{equation}

Since
\[
F_{ilk} = \frac{1}{2} \sum_{\alpha \in R} f'''((\alpha, x)) \alpha_i \alpha_l \alpha_k \qquad (i,l,k \leq n)
\]
where $f'''(u) = \coth (u)$, a simple calculation shows that this condition becomes
\begin{equation}\label{eq7}
\frac{1}{4} \sum_{\alpha \in R, \beta \in R} f''' ((\alpha, x)) f'''((\beta, x))(\alpha,\beta)(\alpha_i \beta_j - \alpha_j \beta_i)(\alpha_l \beta_m - \alpha_m \beta_l) + 2\gamma^2 (\delta_{il} \delta_{jm} - \delta_{jl}\delta_{im})=0
\end{equation}

Considering only positive roots we finally should have 
\begin{equation}\label{eq8}
\sum_{\alpha > 0, \beta > 0} f'''((\alpha, x)) f'''((\beta, x))(\alpha,\beta) (\alpha_i \beta_j - \alpha_j\beta_i)  
(\alpha_l \beta_m - \alpha_m \beta_l)= 2\gamma^2 (\delta_{jl} \delta_{im} - \delta_{il}\delta_{jm})
\end{equation}

For a moment we restrict our attention to the left hand side of the last equality (\ref{eq8}). We split this expression into a partition of pairs of positive roots $\alpha >0, \beta >0$ such that the product $s_{\alpha} s_{\beta}$ of the corresponding Weyl reflections $s_{\alpha}, s_{\beta}$ equals a $w$ in the Weyl group $W$. So we split the sum into
\begin{equation}\label{eq9}
\sum_{w\in W} 
\sum_{
\begin{array}[h]{c}
	\scriptstyle{\alpha >0, \beta >0}\\
	s_{\alpha} s_{\beta} = w
\end{array}} 
f'''((\alpha ,x))f'''((\beta, x)) (\alpha ,\beta)(\alpha_i \beta_j - \alpha_j \beta_i)(\alpha_l \beta_m - \alpha_m \beta_l)
\end{equation}
 
Applying the Dunkl identity (see Matsuo \cite{Mats}, proof of proposition 3.3.1) we see that this sum (\ref{eq9}) equals
\begin{equation}\label{eq10}
\sum_{w\in W} \sum_{
\begin{array}[h]{c}
	\scriptstyle{\alpha >0, \beta >0}\\
	s_{\alpha} s_{\beta} = w
\end{array}} (\alpha ,\beta)(\alpha_i \beta_j - \alpha_j \beta_i)( \alpha_l \beta_m - \alpha_m \beta_l)
\end{equation}

and simplifying this sum again, it equals
\begin{equation}\label{eq11}
\frac14 \sum_{\alpha \in R, \beta \in R} (\alpha, \beta)(\alpha_i \beta_j - \alpha_j \beta_i)(\alpha_l \beta_m - \alpha_m \beta_l)
\end{equation} 

We want to evaluate this last expression. To this end we introduce the homogeneous $4$-form $A$ by
\[
\frac14 \sum_{\alpha \in R, \beta\in R} (\alpha,\beta)((\alpha , x)(\beta,y)-(\alpha,y)(\beta,x))((\alpha , u)(\beta,v)- (\alpha, v)(\beta,u)) = A (x,y;u,v)
\]

Obviously the form $A$ is antisymmetric in $x,y$ and in $u,v$. Moreover it is invariant under the Weyl group $W$ and under permutation of $x,y$ by $u,v$. Consequently a small calculation shows that we necessarily have
\begin{equation}\label{eq12}
A(x,y; u,v) = c((x,u)(y,v)-(x,v)(y,u))
\end{equation}
for some fixed constant $c$. With respect to the Euclidean coordinates  $e_1 , \ldots , e_n$ this means that the expression 
(\ref{eq11}) equals
\[
c (\delta_{il} \delta_{jm} - \delta_{jl} \delta_{im})
\]

Hence the WDVV condition (\ref{eq8}) reduces to $c = -2\gamma^2$. This completes the proof of the theorem.\\

The precise value of $c$ is evaluated with the help of appendix of Bourbaki \cite{Bourb} and is listed in table \ref{tab1}. 
\begin{table}
\begin{tabular}{|c|c|c|c|c|c|c|c|c|}
\hline
    & $A_N$ & $B_N$     & $C_N$       & $D_N$    &$E_6$ & $E_7$ & $E_8$ & $F_4$\\
\hline
 $c$&$2(N+2)$ & $4(2N-3)$ & $8(N+2)$  & $8(N-2)$ & $6$  & 96    & 320   & 30 \\
\hline
\end{tabular}
\caption{The numbers $c$ for each Lie algebra}
\label{tab1}
\end{table}
The numbers listed in this table are in agreement with the results in the paper \cite{Hoev}.

\subsection{Remark}
We may insert a $W$-invariant set of complex numbers, i.e. $k_{w\alpha} = k_{\alpha} (w \in W)$ into the expression $F$ in (\ref{eq2}) in the following way

\begin{equation}\label{eq13}
F(a_1, \ldots, a_n, a_{n+1}) = \frac12 \sum_{\alpha \in R} k_{\alpha} f((\alpha, a)) + \gamma (\frac16 a^3_{n+1} + \frac12 a_{n+1} (a, a))
\end{equation}

The proof of the theorem has to be modified in a rather obvious way. Note that the Dunkl identity remains in force in this case. Of course the value of $\gamma$ has to be modified correspondingly.

\end{document}